\documentclass[preprint,pre,superscriptaddress]{revtex4} %use for single column
\usepackage{tikz}
\usepackage{graphicx}
\usepackage{color} 
\usepackage{transparent}
\usepackage{psfrag}
\usepackage{dcolumn}
\usepackage{amsmath}
\usepackage{amssymb}
\usepackage{latexsym}
 
\usepackage{booktabs}
\usepackage{latexsym}

\usepackage{hyperref}
\begin{document}

%\input{mydefs}
% define short cuts !!
\newcommand{\tikzcircle}[2][red,fill=red]{\tikz[baseline=-0.5ex]\draw[#1,radius=#2] (0,0) circle ;}%
\def\bea{\begin{eqnarray}}
\def\eea{\end{eqnarray}}
\def\beq{\begin{equation}}
\def\eeq{\end{equation}}
\def\f{\frac}
\def\k{\kappa}
\def\e{\epsilon}
\def\ve{\varepsilon}
\def\be{\beta}
\def\D{\Delta}
\def\h{\theta}
\def\t{\tau}
\def\a{\alpha}

\def\cDa{{\cal D}[X]}
\def\cD{{\cal D}[x]}
\def\cL{{\cal L}}
\def\cLo{{\cal L}_0}
\def\cLa{{\cal L}_1}

\def\Re{{\rm Re}}
\def\sj{\sum_{j=1}^2}
\def\rk{\rho^{ (k) }}
\def\rek{\rho^{ (1) }}
\def\cek{C^{ (1) }}
\def\rz{\rho^{ (0) }}
\def\rt{\rho^{ (2) }}
\def\rtb{\bar \rho^{ (2) }}
\def\trk{\tilde\rho^{ (k) }}
\def\trek{\tilde\rho^{ (1) }}
\def\trz{\tilde\rho^{ (0) }}
\def\trt{\tilde\rho^{ (2) }}
\def\r{\rho}
\def\tD{\tilde {D}}

\def\s{\sigma}
\def\kb{k_B}
\def\bF{\bar{\cal F}}
\def\F{{\cal F}}
\def\la{\langle}
\def\ra{\rangle}
\def\nn{\nonumber}
\def\up{\uparrow}
\def\dn{\downarrow}
\def\S{\Sigma}
\def\dg{\dagger}
\def\d{\delta}
\def\p{\partial}
\def\l{\lambda}
\def\L{\Lambda}
\def\G{\Gamma}
\def\o{\Omega}
\def\w{\omega}
\def\g{\gamma}

\def\jv{ {\bf j}}
\def\jr{ {\bf j}_r}
\def\jd{ {\bf j}_d}
\def\jdd{ { j}_d}
\def\noi{\noindent}
\def\a{\alpha}
\def\d{\delta}
\def\p{\partial} 
\def\hf{\frac{1}{2}}

\def\la{\langle}
\def\ra{\rangle}
\def\e{\epsilon}
\def\n{\eta}
\def\g{\gamma}
\def\break#1{\pagebreak \vspace*{#1}}
\def\hf{\frac{1}{2}}

\def\rv{{\bf r}}
\def\pc{\phi_c}
\def\rb{\bar\rho}
\def\ep{\epsilon^\prime}
\def\Rp{R_{\parallel}}
\def\pr{\prime}

\makeatletter
\newcommand{\rmnum}[1]{\romannumeral #1}
\newcommand{\Rmnum}[1]{\expandafter\@slowromancap\romannumeral #1@}
\makeatother

\title{Complex
topologies in phase separated droplets predicted from universal phase diagram}
\author{Amit Kumar}
\email{amit.kumar@weizmann.ac.il}
\affiliation{Dept. of Chemical and Biological Physics, Weizmann Institute of Science, Rehovot, Israel}
%\author{Lucy Brennan}
%\email{amit.kumar@weizmann.ac.il}
%\affiliation{Dept. of Molecular and Cell Biology, University of California, Berkeley, USA}
\author{Gary Karpen}
%\email{amit.kumar@weizmann.ac.il}
\affiliation{Dept. of Molecular and Cell Biology, University of California, Berkeley, USA}
\author{Samuel A. Safran}
\email{Sam.Safran@weizmann.ac.il}
\affiliation{Dept. of Chemical and Biological Physics, Weizmann Institute of Science, Rehovot, Israel}

%\date{\today}

\begin{abstract}
Phase separation of two phase separating solutes in a common solvent can result in mesoscale (micron-sized) droplets with complex topologies of the domains of each solute within each droplet. Such topologies have been
observed in-vitro in systems of chromatin oligomers, biomolecular condensates
and polymeric mixtures. In these systems the solutes phase separate
from the solvent into droplets due to the relatively large free energy gain, which includes the energies
and entropies of mixing with the solvent. Within each droplet, further phase separation can occur
between the two solutes due to an additional free energy difference that promotes their demixing; in some systems, the extent of demixing can be, in some cases, be modulated by an additional component. The minimal free energy topologies are predicted as universal functions of the interfacial tension ratios and fractions of each solute within a droplet.   We compare the predictions with several experimental
systems to estimate the ranges of interfacial tensions. Experimental aspects that may depend on the kinetics or molecular weight variations in the system are also discussed.
\end{abstract}

\maketitle 
%\section{Introduction}
In recent years, there has been an eruption of interest in the study of biomolecular condensates, both in vivo and in vitro~\cite{hyman2014liquid, shin2017liquid, banani2017biomolecular, Berry2018}. The live cell can use the phase separation of some of its components into condensates to organize its biomolecules, such as proteins and nucleic acid polymers~\cite{gibson2019organization,cremer20154d,strom2017phase,larson2017liquid,janssen2018heterochromatin,amiad2021live,Bajpai2021,amiad2023linc,adame2023regulation,kumar2019cross}. Such condensates can serve multiple purposes, including the localization of chemical reactions, stress regulation, gene regulation, DNA repair, etc.  In systems with two or more solutes, each condensates itself can contain a region rich in one of the solutes that coexists with another, rich in the other solute; these coexisting domains can have complex topologies in finite size droplets~\cite{brennan2025hp1a,fisher2020tunable,yanagisawa2014multiple,watanabe2022cell}.  The analytical prediction of these topologies and their intuitive understanding as a function of the ratios of the interfacial tensions of the system components is the focus of this paper. 

Experimental examples of phase separation of two solutes in a common solvent, in which finite droplets with complex topologies of the domains of each solute, include modified (H3K9me3) and unmodified, short oligomers of chromatin~\cite{brennan2025hp1a},  mixtures of poly-L-lysine
(polyK) and poly-L-arginine (polyR)~\cite{fisher2020tunable} and gelatin and polyethylene glycol (PEG) mixtures in aqueous solvent~\cite{watanabe2022cell}.  In the first two examples, the solvent is an aqueous buffer, and there is also an additional component in the system (salt, in the example of polyR and polyK, and the protein HP1a in the chromatin mixture) that tends to  associate with one or both of the solutes. Our theory can apply to such  systems with more than two solutes~\cite{shin2017liquid,yanagisawa2014multiple,gong2024near,motoyoshi2010static,fan2019polymeric,walther2013janus}, if the  equilibrium of the additional component that determines the concentration ratios of this component to the two solutes depends only on the molecular properties of the system and not on their relative volume fractions within the droplets.  This is the case for fixed concentrations of all the components (including the additional one) when the molecular interactions are short-ranged, but the droplet and domain sizes are mesoscopic -- much larger than the interaction range.  One then considers two ``effective'' solutes: the original ones along with their bound fractions of the additional component. This is the approximation that is used here to reduce the analysis of the experimental systems to two effective solutes in a common solvent, where all the interactions and interfacial tensions are``renormalized" by the additional component.

To explain the formation of the domain topologies observed in the condensates of the molecularly different systems mentioned above~\cite{brennan2025hp1a,fisher2020tunable,yanagisawa2014multiple,gong2024near,motoyoshi2010static,fan2019polymeric,walther2013janus}, we consider the generic case of a three component mixture: two ``effective'' solutes (which we term, red and blue) in a common solvent. In the concentration regime above some critical value, the solutes can undergo phase separation into solute-rich and solute-poor regions due to the bulk free energy gain, which includes the ``renormalized"  solute interactions with the solvent and translational entropies~\cite{safran2018statistical, chaikin1995principles}. At asymptotically long enough times, this would lead to one macroscopic phase, the condensate, rich in the two solutes, which coexists with a macroscopic solvent-rich phase.  However, at long but finite times, the condensates are often spherical, micron-sized droplets. Within each droplet, the  red and blue solutes  can themselves be  phase separated due to generally the lower free energy differences between two molecularly similar (but not identical) solutes in an aqueous solvent.  This additional level of phase separation is directly related to the interfacial tensions between the red and blue domains. Within a droplet with a given ratio of the concentrations of the two solutes, the interfacial free energies (product of the appropriate tensions and interfacial areas)  are calculated for each of the topologies we consider.  Comparing those free energies allows us to determine the topology with the lowest free energy for given values of the interfacial tensions~\cite{torza1969coalescence,torza1970three}.  We show that this results in a universal phase diagram that predicts the red-blue topologies within a given droplet as functions of the interfacial tension ratios and the relative red-blue fractions within that droplet.  Since there  are several interfaces in the system (red-solvent, blue-solvent, and red-blue) there can be many such topologies, $e.g.$, crescent, lens, partially engulfed, cap, crescent with cap, lens plus cap, as shown in Fig. 1. When the observed topologies are compared with the theory, the relevant ranges of the interfacial tensions can be extracted.  In addition, by changing the concentration or chemical nature of the additional component, and comparing the observed topology changes with experiment, one can ascertain which tensions are most strongly modulated by the additional component.

Previous models and simulations also predicted some of the topologies (see Fig.~\ref{fig:topo2D_cartoon} ) that we find using a unified framework~\cite{torza1969coalescence,torza1970three,li2017structured}, but to our knowledge, not all; namely, the partially engulfed and cap as well as the cap plus lens, cap plus crescent, shown in the SI.  Fig.~\ref{fig:topo2D_cartoon}  depicts the lens, crescent, engulfed, coating, partially engulfed and coating topologies; these comprise the full set of topologies allowed in the intersection of two circles (in 2d) or two spheres (in 3d).  In addition, instead of numerical results presented in the previous calculations, we display our results as a universal phase diagram of the minimal free energy topologies, as functions of  the tension ratios and volume fractions of red and blue within each droplet; we also predict analytically and explain intuitively the transition between lens and crescent as a function of the tensions and volume fractions. Another recent study focused on the role of hydrodynamics coupled to phase separation~\cite{hester2023fluid,shek2022spontaneous}, but while important for the dynamics, hydrodynamics is probably not a factor in the long-time, mesoscale, steady-state topologies.  In the case of the chromatin, the droplets are viscous so that large-scale hydrodynamic interactions with the droplets is minor. 

The organization of this paper is as follows: First, we present the model, where we discuss the possible set of topologies and the interfacial tensions as a result of phase separation in the three-component system of interest where both solutes tend to phase separate from the solvent and from each other. The Results section presents and explains the universal phase diagram that predicts the minimal free energy topologies as functions of the tension ratios for given ratios of the two solute concentrations within each droplet.  The Discussion discusses the fact that in some experimental systems several topologies can be observed, whereas theory predicts that for each set of tensions and volume fractions, one topology is the lowest free energy.  The multiple topologies observed may hint at polydispersity effects in the solute sizes and/or other kinetic effects.

\section{Model}   

\begin{figure*}[t]
  \centering
  \includegraphics[width=0.9\textwidth]{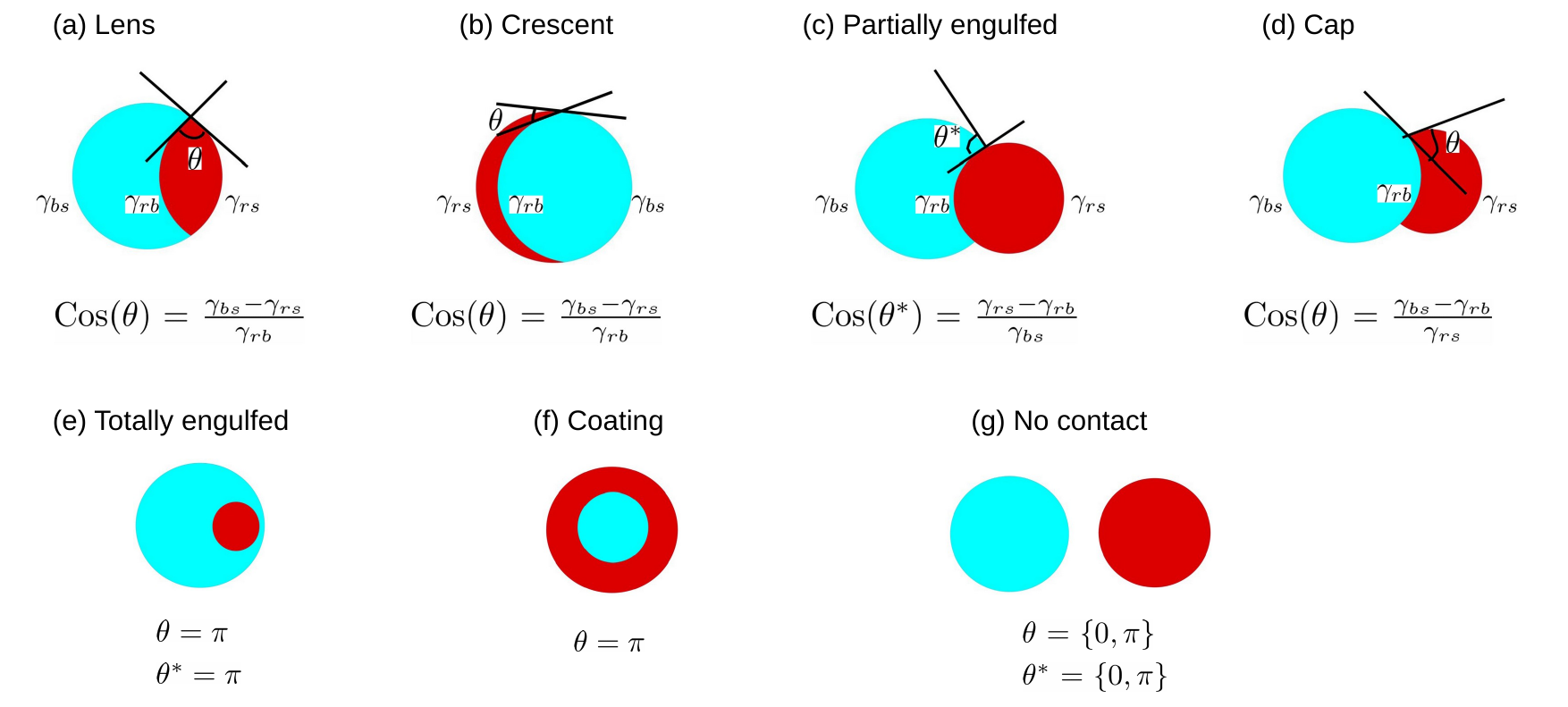}
  \caption{left to right: Lens, crescent, partially engulfed, and cap topologies shown in 2D, respectively. The same topologies occur in 3D from the intersection of two spheres, and the energetics for 3D are discussed in the SI.  The black lines show the tangents to domains with circular (spherical in 3D) curvature at the three interface contacts. The tangents are shown to define the contact angles for the respective topologies at the three-point contact using local force balance (Young's law). The respective contact angles, $\theta$ are defined in the red-rich domain for all the topologies except the partially engulfed case, $\theta^\ast$, where it is defined in the blue-rich domain. Below each topology, we write the respective expression for Young's law as a function of the red-solvent tension, $\gamma_{rs}$, blue-solvent tension, $\gamma_{bs}$ and red-blue tension $\gamma_{rb}$ . The special cases of the topologies in the top panel are shown in the bottom panel: (e) totally engulfed, (f) coating, and (g) no contact topology. These three topologies occur for extreme limits where the contact angles  approach $0$ or $\pi$.}
  \label{fig:topo2D_cartoon}
\end{figure*}

In a mixture of a single solute in a  solvent,  phase separation can occur, for fixed temperature, at concentrations above a given critical value~\cite{safran2018statistical,chaikin1995principles,doi2013soft,pathria2017statistical}. Thermodynamics predicts that demixing of the solution into solute-rich and solute-poor domains occurs when the bulk free energy gain involving the solute molecular interactions and translational entropy is negative. The bulk free energy (whose specification requires a statistical mechanics model for the interactions and entropy~\cite{safran2018statistical,chaikin1995principles,doi2013soft,pathria2017statistical}) determines the chemical potential (derivative of the free energy per unit volume with respect to solute concentration) and osmotic pressure (related to the volume derivative of the free energy).  Thermodynamics, essentially the Second Law, dictates that in equilibrium, two coexisting phases must have the same chemical potential and osmotic pressures. These conditions determine the concentrations of the solute-rich and solute-poor phases.  At the interface of these two phases, there is a gradient of the solute concentration, which involves a free energy cost~\cite{safran2018statistical}. These concepts can be generalized to two solutes in a common solvent.

Here, we consider systems of two solutes that form  condensates where regions of high solute concentration (or either or both types) coexists with regions of high solvent concentration. When there is an additional component which tends to associate with either or both solutes, we defined above ``effective" red and blue solutes with interactions that are ``renormalized'' by the additional component. Furthermore, we consider two effective solutes, red and blue, which themselves tend to phase separate into red-rich and blue-rich phases; the qualifier ``rich'' indicates that a domain of mostly red solute molecules can contain some blue molecules as well as some solvent and vice versa.  Henceforth, for brevity, we shall usually drop the qualifiers of ``effective", ``renormalized", red-rich and blue-rich and refer to red and blue solutes whose interactions lead to phase separation from the solvent into droplets that contain red and blue domains.  This is appropriate for the cases we consider of relatively strong segregation away from any critical points~\cite{safran2018statistical,doi2013soft}.  The various interfacial tensions include those associated with the red-solvent, blue-solvent, and red-blue interfaces; all these values can be renormalized by the additional component as explained above. This is appropriate if we analyze the ensemble of droplets that result from a given average concentration of the solutes and additional component, which is the case in the experiments we discuss.  If those concentrations are varied, the values of the tensions can change.

We note~\cite{safran2018statistical} that net red-blue interaction refers to the direct red-blue molecular interaction relative to the average of the direct red-red and blue-blue interactions. It is this net red-blue interaction that determines the phase diagram and which is related to the interfacial tension between the red and blue domains.  Similar considerations apply to the net red-solvent and net blue-solvent molecular interactions; they are each defined relative to the interactions of those solutes with their own species.  In fact, the extent of phase separation -- the difference in concentration between the red-rich and blue-rich domains within a drop determines the concentration gradient at the interface between those domains.  If this gradient, normalized to the red-rich and blue-rich concentrations within each domain, is small on the molecular scale, the interfacial tension is also small compared to the case where the gradients and tension are large.  This is because the interfacial tension between the red-rich and blue-rich domains, $\g_{rs} \sim \mid \nabla \phi \mid ^2$, where $\nabla \phi$ is the concentration gradient at the interface~\cite{safran2018statistical}.  The positive free energy cost of the interface in phase separating systems, means that the interfacial area between the solute-rich and solute-poor phase will be minimal in equilibrium,  since the free energy is the product of the tension and the interfacial area. For two-component systems (single solute in a solvent), the minimum condition requires that the solute-rich droplets be spheres. For the system of two solutes, the minimum condition requires that each solute domain (red-rich or blue-rich) within a drop be a section of a sphere.  This is because a sphere has the lowest area for a fixed enclosed volume.

Depending on the overall concentrations of the solutes and the temperature, the demixing dynamics of solute molecules and solvent at their late stages typically involves nucleation and growth~\cite{vehkamaki2006classical,mazenko1985instability,kwiatkowski2018phase}.  Small solute-rich droplets nucleate and grow when their size is larger than a critical size, while smaller droplets shrink, with the solute diffusing through the solvent and accruing at the surface of the larger drops.  If the drops are close enough, diffusion and collision of the droplets themselves can also occur. Both these processes predict that the drop radius, $R$ increases with time as: $R(t) \sim t^{1/3}$; this is known as coarsening~\cite{lifshitz1959zh,lifshitz1961kinetics,ratke2002growth,balluffi2005kinetics,bray2002theory}.  The rate of drop coarsening is given by the derivative $dR(t)/dt \sim 1/R(t)^2$; larger drops coarsening significantly more slowly.  Other factors, such as a more complex nature of interaction between solutes, which could result in gel-like, high viscosity solute domains, the presence of surfactants, and long-range electrostatic interactions, can further slow the coarsening process. It is therefore not surprising that coarsening can become very slow for micron-size condensate droplets.  The experiments mentioned above~\cite{brennan2025hp1a,fisher2020tunable,yanagisawa2014multiple,watanabe2022cell}, indeed show micron-scale droplets and not macroscopic phase separation at the experimentally measured time scales.  This motivates our theory which considers \textit{finite-size} droplets that, in general, high concentrations of red and blue solutes (relative to their concentrations in the solvent).  Within these droplets the red and blue molecules can show phase separation into mesoscale domains, sometimes called compartments, that have particular topologies; this is the focus of this paper.

In general, there can be one or more interfaces between the red and blue domains in a drop. In addition, to minimize the interfacial free energy, each domain should be a section of a sphere. Furthermore, since the drop is surrounded by solvent, there is a three-phase line (or point in two dimensions) where the red, blue, and solvent regions coincide (see Fig. ~\ref{fig:topo2D_cartoon}).  For this line to be mechanically stable, the forces tending to minimize the interfacial regions must balance, which leads to Young's law ~\cite{safran2018statistical} relating the interfacial tensions and the contact angle between the local tangents at the three-phase contact line (or point in two-dimensions) as shown in Fig. ~\ref{fig:topo2D_cartoon}. In our system of red and blue solutes in a solvent, the contact angle is determined by the interfacial tensions $\gamma_{rs}, \gamma_{bs}, \gamma_{rb}$, where the subscript $r$ refers to red-rich, $b$ to blue-rich and $s$ to solvent-rich phases.  In general, within a drop, the red and blue domains can exhibit various arrangements of red-blue topologies within a drop, all of which obey  Young's law, which, as a local force balance, acts as a constraint on the overall organization.  However, the most stable organization -- if not hindered by kinetic limitations -- is the one where the overall interfacial energies are minimal.  The interfacial energy is the product of the interfacial tension and the interface area $S_{ij}$ (line length in two-dimensions) for each topology: 
\begin{equation}
E_T = \sum \gamma_{ij} S_{ij}
\label{eq:samenergy1}
\end{equation}
Here, the subscript $T$ denote the seven topologies mentioned in Fig.~\ref{fig:topo2D_cartoon}, and the indices $i$ and $j$ denote the interfaces between  two phases, $i.e.$, red domain (subscript $r$), blue domain (subscript $b$) and solvent (subscript $s$). For example, for the lens topology, the energy expression is given by, 
\begin{equation}
E_{Lens} = S_{bs} \gamma_{bs} + S_{rb} \gamma_{rb} + S_{rs} \gamma_{rs}
\end{equation}.  

The calculation of the interfacial areas (and lengths for the two-dimensional case) are presented in the SI, based on the geometry of two intersecting spheres (or circles in two-dimensions) for the lens, crescent, partially engulfed, and cap topologies shown in Fig. ~\ref{fig:topo2D_cartoon}. For the topologies of lens plus cap or crescent plus cap, the geometry involves three intersecting spheres as shown in the SI. As also shown in Fig. ~\ref{fig:topo2D_cartoon}, limiting cases of the topologies include a blue drop coated by a red shell (or the opposite; see Fig. ~\ref{fig:topo2D_cartoon} $e$ and $d$) or two spherical drops, one red and one blue with no contact between them (Fig. ~\ref{fig:topo2D_cartoon} $g$). 

Minimization of the interfacial energy for the topologies considered leads to one red and one blue domain within a drop (see SI for  explicit calculations for several topologies). The geometrical details associated with a given topology are also determined by the volume fractions of the red and blue domains within a drop such as the: lens, crescent, partially engulfed, and cap  (see Fig.\ref{fig:topo2D_cartoon}). All these topologies can be related to the intersection of two spheres (or circles in two-dimensions), allowing for various sections of these intersections (see SI). The geometry for two spheres, and thus the interfacial areas (lines in two-dimensions) are completely specified by the radii of the two spheres and the distance between their centers.

We have calculated the interfacial energies of the possible topologies and determined the minimum energy topology for a given volume fraction of the red domain or the blue domain. The topology dictates the   interfacial areas (or lengths in two-dimensions, see SI) of the red and blue domains within a drop where their contact angle with the solvent is fixed by Young's law.  We use this framework to predict the topology of lowest interfacial energy for given values of the tensions, $\gamma_{rs}, \gamma_{bs}, \gamma_{rb}$, and the  volume fractions of red and blue in a given drop. 

  We have considered the complete set of topologies which form as the intersection of two spheres and some of those that involve three spheres, for general volume fractions of the red and blue domains. The algebraic and geometric details are presented in the SI. Moreover, we predict a ``phase diagram'' which delineates which topology has the lowest interfacial free energy as a function of the dimensionless ratios of the red-solvent to blue solvent tensions and the red-blue to blue-solvent tensions.  For each volume fraction of red and blue within a drop, this phase diagram is universal, and one can vary the individual tensions in different ways to obtain the same ratios and hence, the same predicted lowest free energy topology.

Before presenting and discussing the predicted ``phase diagram'', we comment on the role of the volume fractions.  In a large range (see SI), the ``phase diagrams'' are similar for different red and blue fractions, except for the transition between the lens and crescent topologies.  Since an overall spherical drop that contains a red lens and a blue crescent (and vice versa), the transitions between these topologies are related and depend more sensitively on the red and blue volume fractions.  In the SI we show that the locus in the tension ``phase diagram'' of the lens to crescent transition is given by a curve in the plane of the ratios $\gamma_1=\gamma_{rs}/\gamma_{bs}$ and $\gamma_2=\gamma_{rb}/\gamma{bs}$ that also depends on the relative red and blue volume fractions.  This relation and its implications is discussed in the next section, Results.

%A few examples of such topologies are given in Fig.~\ref{fig:topo2D_cartoon}.                

%First, at a relatively larger energies compared to translational entropy of the solute molecules, the pahse separation of both type of solutes occur to form solute-rich domains. Usually, at the steady state towards an approach to equilibrium, the rate of droplet coarsening becomers slower at the large times due to scaling $R \sim t^{1/3}$. In addition, various factors such as the complex nature of interaction between solutes, the presence of surfactants and long-range interactions can further debilitate the coarsening process. Consequently, a very slow or an arrested coarsening yield solute-rich droplets at mesoscale. Further, the phase separation between the solute components can occur inside the droplet at the respective lower energies. The phase separation within the droplets leads to the formation of droplets with distinct domains of solute-rich components allowing various conformations of the droplet which can be characterized in distinct topological classes. An example inclde the multiple solute types with differernt solvation embedded in a common solvent.   

\section{Results}

\begin{figure}
  \includegraphics[width=0.7\textwidth]{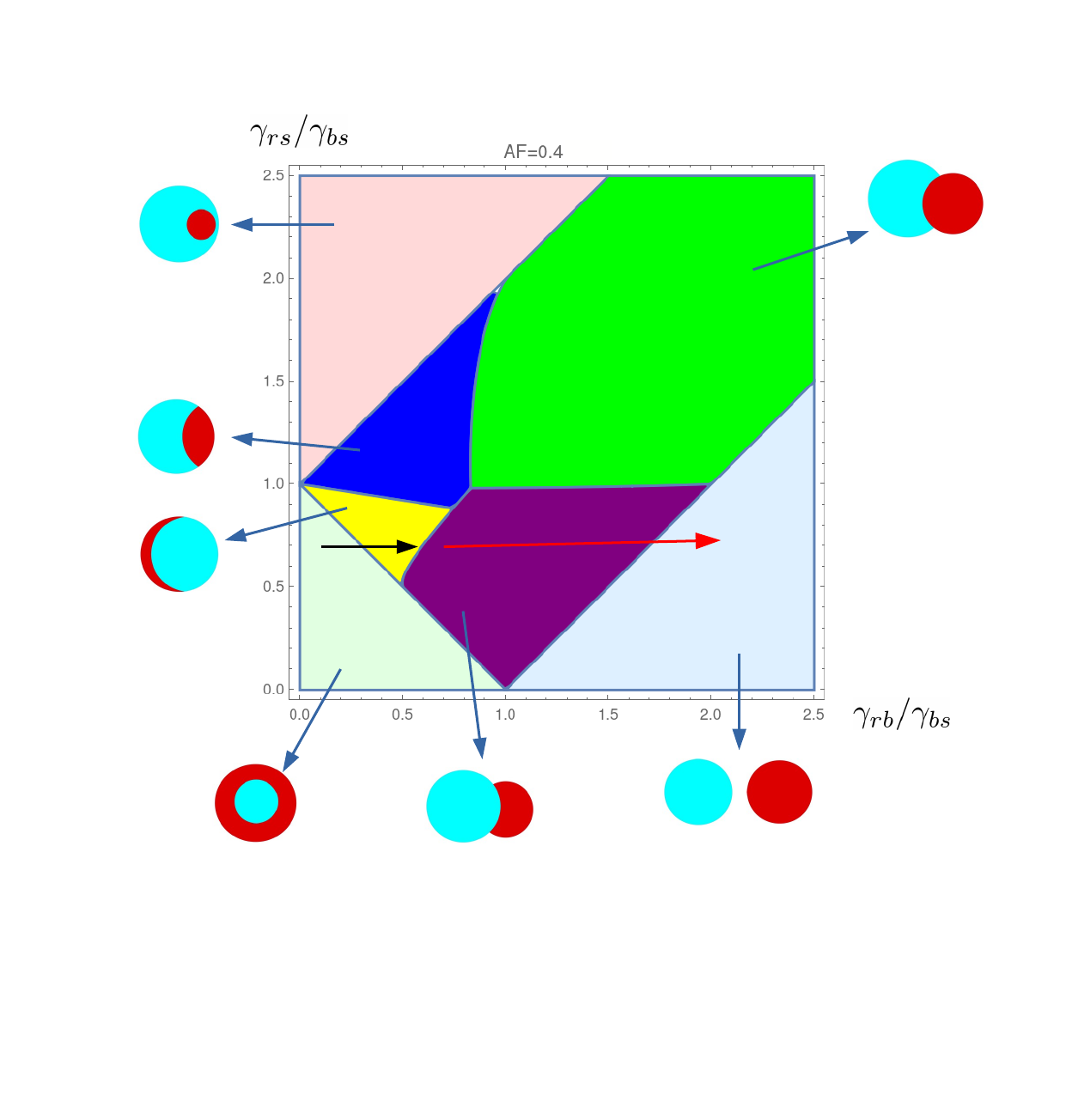}
  \caption{Figure shows the phase diagram of minimum energy topologies in the particular regime of interfacial tensions $\gamma_{rs}$, $\gamma_{bs}$ and $\gamma_{rb}$. X-axis and Y-axis represents the interfacial tension ratios $\gamma_1 = \gamma_{rs}/\gamma_{bs}$, and $\gamma_2 = \gamma_{rb}/\gamma_{bs}$ respectively. Each color represents the minimum energy topology among all the topologies mentioned in the phase diagram. The minimum energy topology in each region is also shown for convenience. The red arrow, followed by the black arrow, shows the axis of increasing $\g_2$ and the change in associated topologies.}
  \label{fig:phase_diagram1}
\end{figure}

To estimate the minimum energy topology, we calculated the interfacial energy defined above of all the droplets that have the same bulk energy, which is proportional to the volume fractions of the red and blue domains.  Thus, our final ``phase diagrams'' for the topologies as functions of the tension ratios $\gamma_1=\gamma_{rs}/\gamma_{bs}$ and $\gamma_2=\gamma_{rb}/\gamma{bs}$  are each calculated for a given ratio of the red and blue volumes (areas in two-dimensions).   The Supplementary Information (SI) contains the details of these calculations. We calculate the line lengths of various circular interfaces (area for 3D spherical interfaces) for all the topologies determined by radii of intersecting circles and the separation of their respective centers. The line lengths are multiplied by the respective interfacial tension and added to obtain the total surface energy of a given topology.

Our results are summarized in a universal ``phase diagram'' in the plane of $\gamma_1$ and $\gamma_2$ which are ratios of the tensions, as shown in Fig.~\ref{fig:phase_diagram1}.  This is calculated for the simpler geometry of intersecting circles, although this is very similar in three-dimensions.  The topologies indicated there are the lowest interfacial energy topologies for those considered and exhaust all the topologies formed by the intersection of two circles (or two spheres in three-dimensions). We have explicitly calculated the transitions between the completely engulfed, coating, lens, and crescent topologies in three-dimensions and found that they are very similar to the transitions found in two-dimensions (see SI).  In the SI, we also include the cases of a lens plus cap and crescent plus cap and compare those to each other and to the other topologies shown in Fig.~\ref{fig:phase_diagram1}.  However, this results in only minor changes for the parts of the  ``phase diagram'' relevant to the experiments.

For example, comparison of the interfacial energies shows that lens topology has the minimal energy of all the topologies considered in the region of the ``phase diagram'' Fig.~\ref{fig:phase_diagram1} indicated in the figure. This occurs for relatively small values of $\gamma_2$ and moderate values of $\gamma_1$ and shares a boundary with the region where crescent topology has minimum energy as shown in the figure.  A red lens has convex curvature vis a vis the blue domain of the drop (see Fig. \ref{fig:topo2D_cartoon} a), while a red crescent has convex curvature.  Those two curvatures are equal when the interface separating red and blue domains is flat.  In that case, the crescent and lens topologies are the same and have the same interfacial free energy (for a given volume fraction).  The transition between them is delineated by the line between the crescent and lens in the ``phase diagram'' as a function of $\gamma_1$ and $\gamma_2$. This criterion depends on the volume fractions (or area fractions in two-dimensions) of the red and blue domains as derived in the SI.  For example, in two-dimensions, the transition as a function of the red area fraction, $A_r$ is 
\begin{equation}
 A_r = (\theta - \cos(\theta) \sin(\theta))/\pi,
\end{equation} where $\theta$ is the contact angle, whose dependence on the tensions is the same for the lens and the crescent and is given in Fig.~\ref{fig:topo2D_cartoon} by Young's law: $\cos(\theta) = (1-\gamma_1)/\gamma_2$.

We note that an increase in $\gamma_{rs}$ away from the lens to crescent transition line in the ($\gamma_2$, $\gamma_1$) plane results in an increment in the contact angle due to the change in curvature towards the lens topology. This can also be achieved by decreasing $\g_{bs}$. At smaller values of $\gamma_{bs}$ or larger values of$\gamma_{rs}$, it is energetically more favorable for the blue domain to have a larger interface with the solvent so that the red domain forms  a crescent at the same volume fraction. Along the $\gamma_2$ axis, the transition line between the lens and crescent topologies is given by $\gamma_1=1$ for equal red and blue volumes (areas in two dimensions), $A_r=1/2$; the slope of that line decreases in the ($\gamma_2$, $\gamma_1$) plane as the fraction of the red domain decreases. This implies that for fractions less than 1/2, the lens region in the      ``phase diagram'' is larger than the crescent region. As  $\gamma_{rb}$ is increased, it is more favorable energetically to have a smaller interface between red and blue domains; this results in the lens topology having lower energy than the crescent. 

As the ratio $\g_1$ of the red-solvent to blue-solvent tensions increases further,  the energy required to form red-solvent interface increases, and the lowest interfacial energy topology is one that maximizes the interface of the blue domain with the solvent.  Thus, the ``phase diagram'' in this regime shows that the engulfed topology is lower energy than the lens topology.  In the engulfed topology,  a red spherical drop is completely encapsulated within the blue domain; only the blue domain has an interface with the solvent.  The other extreme occurs as $\g_1$ is decreased to small values, which makes it energetically preferable to maximize the interface between the red domain and solvent (given the volume fraction constraints). Consequently, the red domain encapsulates the blue domain, forming a coating rather than the crescent topology. In short,  there is a smooth transition from lens to engulfed topology as $\gamma_1$ is increased and from crescent to coating as $\gamma_1$ is decreased.  

We now discuss the other topologies shown in the ``phase diagram'', using two-dimensional language of circles; the geometry in three dimensions (of spheres) is very similar.  The topologies shown in Fig.\ref{fig:topo2D_cartoon} (c), partially engulfed, and (d) cap topology, are the possible intersections of one complete red circular domain, and one blue crescent domain.  This leads to a major difference from lens and crescent topologies, which both fit in one circular domain,  while (c) and (d) require two circles for their definitions. This gives these topologies additional freedom, so long as the contact angle constraints are obeyed (see Fig. 1), for fixed tension ratios. Consequently, the transition lines between the two topologies ((c) and (d)) are solely determined by the contact angles and hence the tension ratios, but not the area fractions of blue and red.  In the case of the partially engulfed topology, a red circular domain forms an interface with the blue crescent domain. This is the only topology in which the contact angle is defined within the blue (e..g., majority in Fig. 1) domain, and the region in the ``phase diagram'' where it is the minimal interfacial tension indicated in  Fig.~\ref{fig:phase_diagram1}. The partially engulfed topology is lowest energy for relatively large values of $\g_1$ and $\g_2$ and is almost symmetric around a diagonal line in the ($\gamma_2$, $\gamma_1$) plane. The partially engulfed region of the ``phase diagram'' shares a boundary with the lens region with a slightly curved transition line.  For smaller volume fractions of the (minority) red domain, a narrow transition region also occurs between partially engulfed and crescent (see SI). The transition between the two topologies, $i.e.$, lens to partially engulfed, and crescent to partially engulfed, is probably discontinuous as it involves abrupt changes in topology.  

The partially engulfed topology also shares a boundary with the cap topology, as shown in the ``phase diagram''. The transition line between them follows the line $\g_1$ = 1, and is constrained by a geometrical requirement on the contact angle. When $\g_1$ = 1, Young's law is the same for both topologies (cap and partially engulfed), $\cos(\theta^\ast) = \cos(\theta) = 1 - \g_2$.         

In addition to the topologies mentioned in Fig.~\ref{fig:topo2D_cartoon}, we considered two additional topologies: lens plus cap and  crescent plus cap. Both topologies require for their definition, the intersection of \textit{three} circles (see SI). Consequently, the three interfaces have different magnitudes of curvatures in general, contrary to previous cases where at least the two interfaces were a part of the same circle having the same curvature.  For the lens plus cap topology, the drop (which is no longer spherical) includes a single spherical drop with a minority phase red lens (and majority phase blue crescent), together with an additional red cap. 

 For the crescent plus cap topology, the drop (which is no longer spherical) includes a single spherical drop with a minority phase red crescent (and majority phase blue lens) together with an additional red cap.  At the three-phase contact point, the geometrical construction of tangents to the domains leads to two contact angles, as shown in the SI. Young's law is different for the two contact angles and is a nonlinear function of the tensions (see SI). The regions  in the ``phase diagram'' where these more complex topologies have lower energy than the others is shown in the SI.

\section{Discussion}   
In this paper, we analyzed the phase separation of two solutes (red and blue) in a common solvent to explain the observed topologies in experiments. We focused on a system where the interactions are such that both solute phase separate from the solvent~\cite{safran2018statistical,doi2013soft}.  Although the very long time behavior would be one of macroscopic phase separation, at finite times, the phase separation occurs in spherical, (circular in 2D) solute-rich droplets. The interactions of the red and blue solutes are such that the two solutes themselves phase separate into red and blue domains within each droplet. As a result, the interplay of interfacial tensions (red-solvent, blue-solvent, and red-blue) can lead to many possible topologies (see Fig. 1) where the red and blue domains within each sphere are formed from sections of spheres. Since spheres are the shape of minimal area that enclose a finite volume,  this geometry minimizes the interfacial energies. We showed that the minimal interfacial free energy topology is a function of interfacial tensions ratios but also depends on  the relative volume fractions (area fractions in 2D) of the red and blue domains within the droplet.

From this theoretical interfacial energy minimization, we predicted a shape phase diagram of the  minimum energy topologies shown in Fig. 2 (and for different area and volume fractions in the SI) that is universal since it only depends  on the tension ratios and relative volume fractions.  Since the bulk energies of each domain must be the same for the comparison of the interfacial free energies (see SI), we compare  the interfacial free energies per unit volume (or area in 2D) for each topology for the same relative volume fractions and same interfacial tension values.   In particular,  the transition between the crescent and the lens topologies are strongly dependent on the area or volume fractions as we showed analytically (see SI for details). This is because in the case of the lens and crescent topologies, the red and blue subdomains are constrained to be accommodated within a spherical (circular in 2D) droplet. That constraint  leads to an explicit dependence on the relative amounts of red and blue domains within the droplet. The transitions between the other topologies ($e.g.$ cap and partially engulfed) are mostly determined by the geometry, $i.e.$, the contact angle, since these  topologies provide more freedom for the red-rich and blue-rich domains to minimize the total interfacial free energy. However, the transition lines that share a boundary with either the lens or crescent topology also show dependence on the area or volume fraction of the subdomains for the reason explained above. 

If one follows the black and red arrows in the universal phase diagram of Fig.~\ref{fig:phase_diagram1}, there is a succession of topological transitions: coating to crescent and cap to no contact. This path can correspond to an increase in the red-blue interfacial tension, $\g_{rb}$, which would corresponding to increasing the net (see the Model section for its definition), red-blue interaction. In the experimental system of methylated and unmethylated chromatin condensates, the increase in the net red-blue interaction is controlled by variants of the additional component HP1a which is subsumed in the ``effective" red and blue interactions and tensions as explained in the Introduction section. This is schematically depicted in Fig. 5 of the experimental paper~\cite{brennan2025hp1a} and in the increased red-blue contrast seen in Fig. 5A for the different HP1a variants. In another experimental example of condensates containing both poly-L-lysine (polyK), poly-L-arginine (polyR) sequences, in the presence of nucleobase and the salt uridine-5’-triphosphate trisodium salt (UTP), the change in the ``effective" interactions of the two components via  variation of salt concentration, leads to changes in topologies from coating and crescent as shown in the middle panel of  Fig. 6A of ~\cite{fisher2020tunable} at the longest times. Similar trends are observed for a blend of synthetic polymers (PEG and gelatin~\cite{yanagisawa2014multiple}, PEG and dextran~\cite{watanabe2022cell}), where different topologies are observed as the chemical composition (Fig. 1C  of Ref.~\cite{yanagisawa2014multiple}, Fig.1D of Ref.~\cite{watanabe2022cell}) is varied.

Our universal phase diagram defines a region in the plane of the interfacial tension ratios (the axes of Fig. 2) where a given observed topology has the minimal interfacial free energy; for brevity, we call this the ``topology region". In microscopy experiments such as those involving chromatin oligomers~\cite{brennan2025hp1a},  condensates of poly-L-Lysin and poly-R-argenin ~\cite{fisher2020tunable}, and synthetic polymer mixtures~\cite{yanagisawa2014multiple,watanabe2022cell}, etc., the contact angle and the relative area  fractions (volume fractions for 3D images) of the red and blue domains can also be measured. The measured contact angle combined with Young's law for the topology considered (see Fig. 1) determines a line in the plane of the relative interfacial tensions.  Where this line occurs within the topology region delineates the possible values of the ratios of the interfacial tension for a given topology.  Thus, our phase diagram, considerably limits these ratios compared to the entire line of tension ratios determined by the measured contact angle. 

Our theory is idealized and mean-field in nature.  We assume that the system consists of identical red and blue solutes where the red-red, blue-blue, blue-red, red-solvent, blue-solvent interactions  each have well-defined values that do not deviate within the system. In addition, we assume that the concentrations of solutes in each domain (red-rich, blue-rich and solvent) are determined by the usual thermodynamic conditions of equal chemical potential. This  leads to well-defined interfacial tension values in the system.  In true thermodynamic equilibrium, the micron-scale droplets would coarsen to result in macroscopic phase separation of the red-rich, blue-rich and solvent domains.
We next discuss how the experimental conditions can differ from the idealization of our theoretical model. 

\textit{Coarsening kinetics:}In the experiments,  coarsening of the drop size distribution can saturate after a long but finite time of order of an hour ~\cite{brennan2025hp1a,fisher2020tunable,yanagisawa2014multiple,watanabe2022cell}. The observed slow growth of large drops can be attributed to either Ostwald ripening or droplet diffusion~\cite{voorhees1992ostwald,taylor1998ostwald}.  In some cases, the solutes within a droplet may form a highly viscous gel and either topological entanglements of polymers or cross-linking can kinetically prevent the merger of colliding drops. Further, in LLPS the drops are known to have ultra-low interfacial tensions (4-5 orders of magnitude lower than typical hydrophilic-hydrophobic interfaces) with the aqueous buffer~\cite{hyman2014liquid,caragine2018surface,law2023bending,kumar2023fluctuations}, which would lead to very slow coarsening, since the tension is a driving force behind coarsening. The presence of a surface layer of the same polymer but with a possibly different conformations might be responsible for the ultra-low tension~\cite{golani2025mesoscale} and stabilize the droplets similar to emulsions.  In any case, the topologies considered here and in the experimental systems are indeed relevant to the observed mesoscale droplets formed on the scale of hours whose much longer time properties are so far unknown.

\textit{Additional components:} To assess the tension effects that govern the droplet topologies, one can correlate the tension magnitude with the difference in the concentrations of the two solutes (red and blue) in each of the coexisting phases. This is because the tension itself originates from this phase separation~\cite{safran2018statistical}, which can be ``weak'' (the concentration difference of red and blue solutes between the two phases is small) or ``strong'' (large concentration difference, with almost all red solutes in one phase and blue solutes in the other). Weak and strong correlate with the strength of the red-blue interactions relative to the temperature. As pointed out in the Introduction, these are  ``effective" interactions, since they can be modulated by additional components.  In addition, the net (see Introduction for its definition) red-blue  interaction, and hence tension, is related to the difference between the direct, ``effective" red-blue interactions and  the direct, ``effective'' red-red, and blue-blue interactions~\cite{safran2018statistical}.  Thus, changes in an additional component can change any or all of these and hence the associated tensions. For example, the experiments on solutions of methylated and unmodified chromatin oligomers are conducted in the presence of the protein HP1a.  The ``effective" red and blue solutes would then be the methylated oligomer plus its associated HP1a and the unmodified oligomer with its associated HP1a. Variants of the HP1a protein thus tune the ``effective" interactions between the red and blue solutes and the solvent~\cite{brennan2025hp1a}. Interventions were also used in the experiments on biomolecular condensates of polyR and polyK in Ref.~\cite{fisher2020tunable}.  There, the additional components were nucleobase and salts.  In terms of the theory, modification by the additional component of the net, ``effective" red-blue tension can occur, for example, by the additional component increasing  the direct ``effective" red-red attractions, which then give rise to a smaller value of the net, ``effective" red-blue interactions and the associated tension. It can of course, also modulate the direct ``effective'' red-blue interactions.

We explained above that the extent of the phase separation of the two solutes can be ``strong" or ``weak"; in the latter case, the concentration difference of the two solutes in the two coexisting phases within a droplet is relatively small. In experiments, the extent of phase separation can be inferred from the color contrast between the red and blue domains. For example, in the experimental case of chromatin oligomers in the presence of HP1a, the variations in the mutations of the protein HP1a is used to change the effective interaction between the methylated (red solute) and unmodified chromatin (blue solute). Consequently, the condensates show varying contrast between subdomains (methylated and unmodified) with the different mutants present as it can be seen in Fig.5(A) of Ref.~\cite{brennan2025hp1a}.   Apart from the color contrast, changes in the topology can also be used to infer the nature of changes in the  interactions between the ``effecive" red and blue solutes. This effect is clearly displayed in other experimental cases (Fig. 6A of Ref.~\cite{fisher2020tunable}), in addition to the chromatin condensates~\cite{brennan2025hp1a}.

\textit{Kinetic limitations on red and blue domain topologies:} While our idealized theory predicts the minimal free energy topology of the red and blue domains within a single droplet, we should bear in mind that stronger interactions at the molecular level may have important kinetic implications. In general, in the absence of such limitations, only one red and one blue domain are expected in each droplet. However, this may not always be the case experimentally.  The presence of a micron-scale drop probably resulted from the collisions of many smaller drops, and as a result of these collisions, a micron scale droplet at some early time may have several red and several blue regions, corresponding to those of the smaller drops. The interfacial energy of such a configuration is not minimal, and given sufficient kinetics, the smaller domains are expected to merge, resulting in one red and one blue region per drop. However, the kinetic barriers to separating molecules once they have associated may be too high for equilibration to occur, and in particular strong interactions among the solutes may lead to such kinetic barriers. In addition, if the individual droplets are not dilute -- as appears from the experimental pictures~\cite{brennan2025hp1a,fisher2020tunable} -- collisions of large droplets may occur, which can lead to additional kinetic barriers  which cannot be overcome over the time of the experiment.  In some cases, such kinetic limitations might be overcome by raising the temperature (annealing) and then lowering it slowly  the system is prepared. Since little is known about the ``best'' absolute cooling rate, one must vary those rates in a systematic manner to ascertain how slowly the system must be cooled to reach the equilibrium topology within a given droplet. One might also study systems in which large (micron-scale) droplets are well-separated; in such dilute systems, collisions and the attendant kinetic effects for strong interactions may be less probable.  Finally, strong crosslinking or entanglement of polymeric solutes can create highly viscous, gel-like domains whose equilibration might not occur on the experimental time scales.  Again, this is most likely for the systems with strong red-blue interactions, where indeed, the structures are more inhomogeneous compared to the theoretical predictions as observed in the bottom two panels of Fig. 5A~\cite{brennan2025hp1a}.

\textit{Multiple topologies in the same system:} The observation of more than one topology  for same experimental setup may suggest that the compositions and interfacial tensions are not identical in all the droplets.  A given experimental system may include drops with somewhat different values of the tension ratios, and hence from our phase diagram, several topologies -- some occurring frequently, and some rarely.
Indeed the experiments show in some cases, two or more topologies for the same experimental setup (e.g., coating and crescent, crescent and cap, and cap and no-contact) [see Fig.5A, third row, first column coating and crescent, fourth row, first column crescent and cap of Ref.~\cite{brennan2025hp1a}, Fig. 6A coating and crescent of Ref.~\cite{fisher2020tunable}, Fig. 1C  of Ref.~\cite{yanagisawa2014multiple}, and Fig.1D of Ref.~\cite{watanabe2022cell}].  This is in contrast to our idealized theory, which predicts a single topology in a given droplet as the minimal energy  state, for fixed values of all the molecular interactions.  We note that the two observed topologies (e.g., coating and crescent) for each experimental setup are adjacent in the theoretical phase diagram; their tension ratios are relatively close.  It may be that this occurs because of variations in the molecular interactions of the components of a given drop.  For polymeric solutes, this can  arise from a dispersion of chain lengths which has been quantified (in terms of the radius of gyration) in Fig. 3F of Ref.~\cite{brennan2025hp1a}.  Different drops may contain polymers of different lengths and thus different red-blue interactions. There is a similar issue of possibly different contact angles in the same system. If those angles can be accurately measured by image processing, they can provide information on the range of polymer chain lengths or other inhomogeneities in the system.  Phase separation of polymers can be quite sensitive to the chain length~\cite{de1979scaling,doi2013soft}, since for large chain lengths, the translational entropy becomes negligible relative to the intermolecular interactions. 

There is therefore some scope for future experiments that may allow a more accurate comparison of the theory with the observed  topologies and contact angles.  While  extremely slow cooling rates may be impractical, one can test if slower and slower cooling rates give rise to a more homogeneous distribution of the observed topologies and contact angles. Similarly, while some polydispersity of chain lengths cannot be avoided, experiments may investigate whether a trend towards less polydisperse systems result in more homogeneous results. Such experiments will allow for a more accurate appraisal of the two interfacial tension ratios in the red, blue, solvent system based on our theory, when combined with the contact angle measurements, provides an estimate of the range of those in a restricted region dictated by the topology of the domains.

The authors are grateful to Lucy Brennan for many valuable discussions.  We  also thank Arash Nikoubashman and Miho Yanagisawa for their comments.  SAS and GHK are grateful to the Volkswagen foundation for its support in grant 197/98.  SAS, who holds the Fern and Manfred Steinfeld chair, thanks the historic generosity of the Perlman Family Foundation.

\bibliographystyle{apalike}
\bibliography{main}

\end{document}